\documentclass[twocolumn,,amsmath,amssymb]{revtex4-1}

\usepackage{graphicx}
\usepackage{dcolumn}
\usepackage{bm}
\newcommand{\etal}{\emph{et al.}}
\newcommand{\be}{\begin{equation}}
\newcommand{\ee}{\end{equation}}
\newcommand{\bfig}{\begin{figure}}
\newcommand{\efig}{\end{figure}}

\usepackage{lineno}
\usepackage{lipsum}

\begin{document}      

\title{Oscillations of the thermal conductivity observed in the spin-liquid state of $\alpha$-RuCl$_3$
} 

\author{Peter Czajka$^{1,*}$}
\author{Tong Gao$^{1,*}$}
\author{Max Hirschberger$^{1,\dagger}$}
\author{Paula Lampen-Kelley$^{2,3}$}
\author{Arnab Banerjee$^{4,\ddagger}$}
\author{Jiaqiang Yan$^3$}
\author{David G. Mandrus$^{2,3}$}
\author{Stephen E. Nagler$^{4}$}
\author{N. P. Ong$^{1,\S}$}
\affiliation{
{$^1$Department of Physics, Princeton University, Princeton, NJ 08544, USA}\\
{$^2$Department of Materials Science and Engineering, University of Tennessee, Knoxville, Tennessee 37996, USA} \\
{$^3$Materials Science and Technology Division, and $^4$Neutron Scattering Division, Oak Ridge National Laboratory, Oak Ridge, Tennessee 37831, USA}
}

\date{\today}      

\maketitle     

{\bf In the class of materials called spin liquids, a magnetically ordered state cannot be attained even at milliKelvin temperatures because of conflicting constraints on each spin (for e.g. from geometric or exchange frustration). The resulting quantum spin-liquid (QSL) state is currently of intense interest because it exhibits novel excitations as well as wave-function entanglement. The layered insulator $\alpha$-RuCl$_3$ orders as a zigzag antiferromagnet below $\sim$7 K in zero magnetic field. The zigzag order is destroyed when a magnetic field $\bf H$ is applied parallel to the zigzag axis a. Within the field interval (7.3, 11) Tesla, there is growing evidence that a QSL state exists. Here we report the observation of oscillations in its thermal conductivity below 4 K. The oscillation amplitude is very large within the interval (7.3, 11) T and strongly suppressed on either side. Paradoxically, the oscillations are periodic in 1/\emph{H}, analogous to quantum oscillations in metals, even though $\alpha$-RuCl$_3$ is an excellent insulator with a gap of 1.9 eV. By tilting $\bf H$ out of the plane, we find that the oscillation period is determined by the in-plane component $H_a$. As the temperature is raised above 0.5 K, the oscillation amplitude decreases exponentially. The decrease anticorrelates with the emergence above $\sim$2 K of an anomalous planar thermal Hall conductivity measured with $\bf H\parallel a$. To exclude extrinsic artifacts, we carried out several tests. The implications of the oscillations are discussed. 
}

\vspace{3mm}\noindent
The Quantum Spin Liquid (QSL), first described by Anderson~\cite{Anderson}, is an exotic state of matter in which the spin wave functions are highly entangled, but long-range magnetic order is absent~\cite{Ng,Savary,WenQFT}. The Kitaev honeycomb model Hamiltonian $H_K$ for a class of spin liquids has attracted intense interest because the exact solution of its ground state in zero magnetic field features anyonic excitations that are Majorana fermions and $Z_2$ vortices~\cite{Kitaev}. 

\begin{figure*}[t]
\includegraphics[width=18 cm]{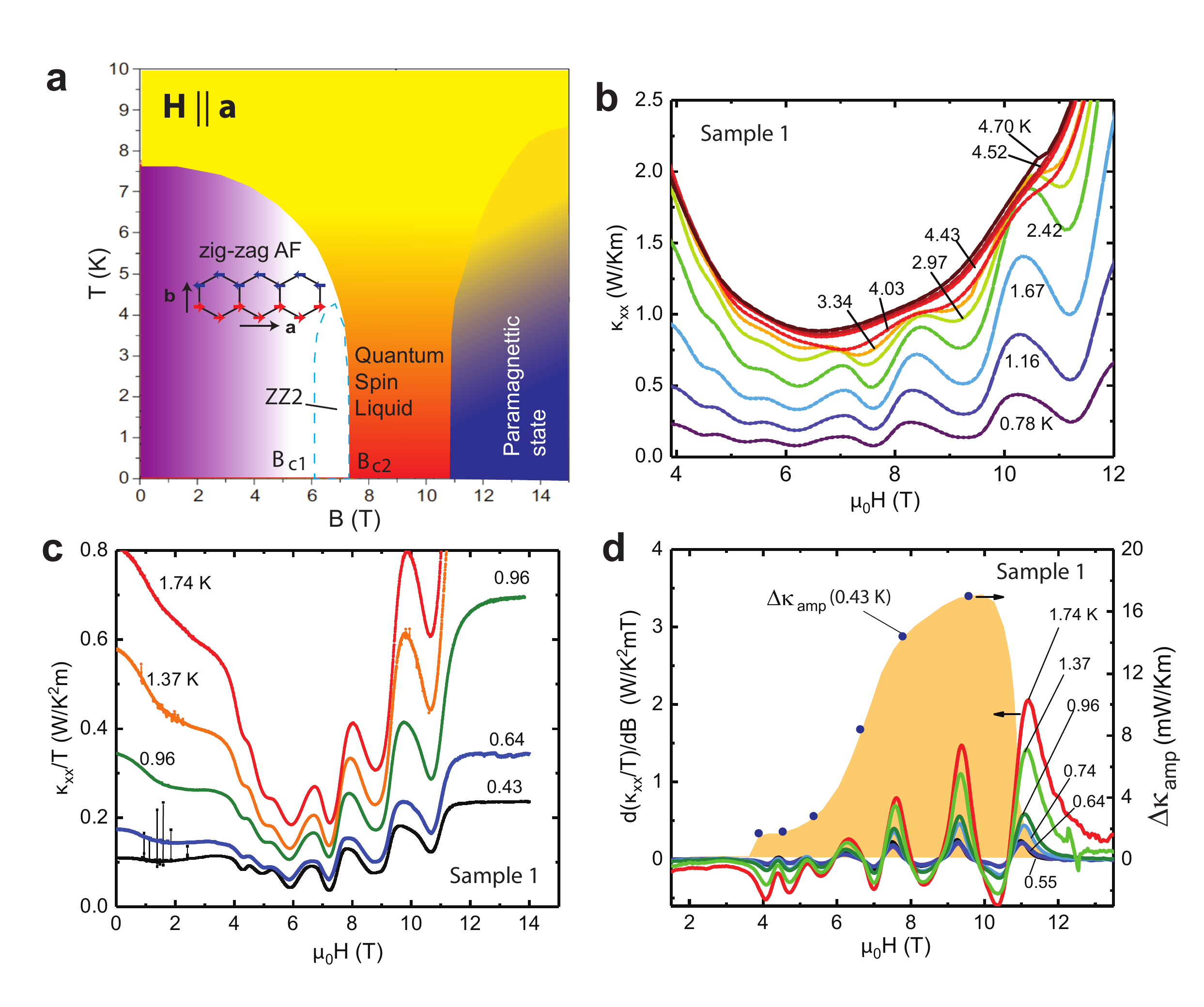}
\caption{\label{fig1} 
}
\end{figure*}

The magnetic insulator $\alpha$-RuCl$_3$ is proximate to the Kitaev honeycomb model~\cite{Plumb,Sears1,Banerjee,Sears2,Wang,Leahy,Banerjee2,Hentrich,Lampen1,Lampen2,Balz}.
Interaction between the spins on Ru ions are described by the Kitaev exchange terms, e.g. $K_{X} \sigma^{X}_i\sigma^{X}_j$ where $X,Y$ and $Z$ define the spin axes~\cite{Khaliullin,Chun}. Additional exchange terms $\Gamma$ and $\Gamma'$~\cite{KimKee} stabilize a zig-zag  antiferromagnetic state below 7 K when the magnetic field $\bf H$=0 (Fig. \ref{fig1}a, inset). 

In a field $\bf H\parallel a$, the zig-zag state is suppressed when $H$ exceeds the critical field $H_C$ = 7.3 T. Within the field interval (7.3, 11) T, experiments ~\cite{Banerjee,Balz} reveal a magnetically disordered state, identified as a quantum spin liquid (QSL) in which magnons give way to very broad modes~\cite{Wang,Balz}. Above 11 T, the local moments are partially field-polarized. Interest was heightened by a report~\cite{Kasahara} that the thermal Hall conductivity $\kappa_{xy}$ is quantized within a 2-Kelvin window (from 3.3 to 5.5 K).

Here we report measurements of both $\kappa_{xy}$ and the thermal conductivity $\kappa_{xx}$ to temperatures $T\sim$ 0.4 K. We have uncovered oscillations in $\kappa_{xx}$ with amplitudes that are strongly peaked when $H$ lies in the interval (7.3, 11) T. Tilting $\bf H$ reveals that the oscillation period is determined by the in-plane component $H_a$. We note that, despite the similarity to Shubnikov de Haas (SdH) oscillations in metals, the free-carrier population at 1 K is exponentially suppressed by the large gap of 1.9 eV~\cite{Sinn}.

Above $\sim$2 K, we recorded using the stepped-field method both the thermal resistivity $\lambda_{xx}$ and thermal Hall resistivity $\lambda_{yx}$ as $H$ was slowly varied at fixed $T$ (Secs. A and B of Methods). As seen in Fig. \ref{fig1}b (for Sample 1 with $\bf H\parallel a$), strong oscillations emerge in $\kappa_{xx}(H)$.  Below $\sim$2 K, the data were recorded continuously as well with the stepped-field method. The data, plotted as $\kappa_{xx}/T$ in Fig. \ref{fig1}c, show that the oscillation amplitudes continue to grow until they comprise 30-60$\%$ of $\kappa_{xx}$ at 0.43 K. At $\sim$11.5 T, $\kappa_{xx}$ displays a step-increase to a flat plateau. In the high-field partially polarized state, where $\kappa_{xx}$ is dominated by the phonon conductivity $\kappa^{ph}$, oscillations are rigorously absent (see below and Sec. \ref{phonons} in Methods). Similar curves are observed in Sample 3 (Fig. \ref{figS3}a in Methods).

The oscillation amplitudes are strongly peaked in the QSL state. To extract the amplitude, we first determined the smooth background curve $\kappa_{bg}(T,H)$ threading the midpoints between adjacent extrema (Fig. \ref{figS2}d in Methods). The oscillatory component, defined as $\Delta\kappa = \kappa_{xx}-\kappa_{bg}$ (Fig. \ref{figS2}c in Methods), allowed accurate determination of the amplitude $\Delta\kappa_{amp}$, which we plot in Fig. \ref{fig1}d for Sample 1 (solid circles). Above 6 T, $\Delta\kappa_{amp}$ rises steeply to peak at 9.6 T, followed by an abrupt collapse to zero above 11 T. Below 6 T, a weak remnant ``tail'' survives to 4 T in a mixed state in which small QSL regions coexist with the zig-zag state (we note that 4 T is roughly where the averaged zig-zag Bragg intensity begins to weaken with $H$~\cite{Banerjee2}). By its profile, $\Delta\kappa_{amp}$ is largest within the field interval (7.3, 11.5 T) of the QSL state. The profile in Sample 3 is similar (Fig. \ref{figS3}a of Methods). A fourth sample 4 did not exhibit oscillations with $\bf H$ tilted at 45$\circ$ to $\bf a$ (Table in Methods).

\begin{figure*}[t]
\includegraphics[width=18 cm]{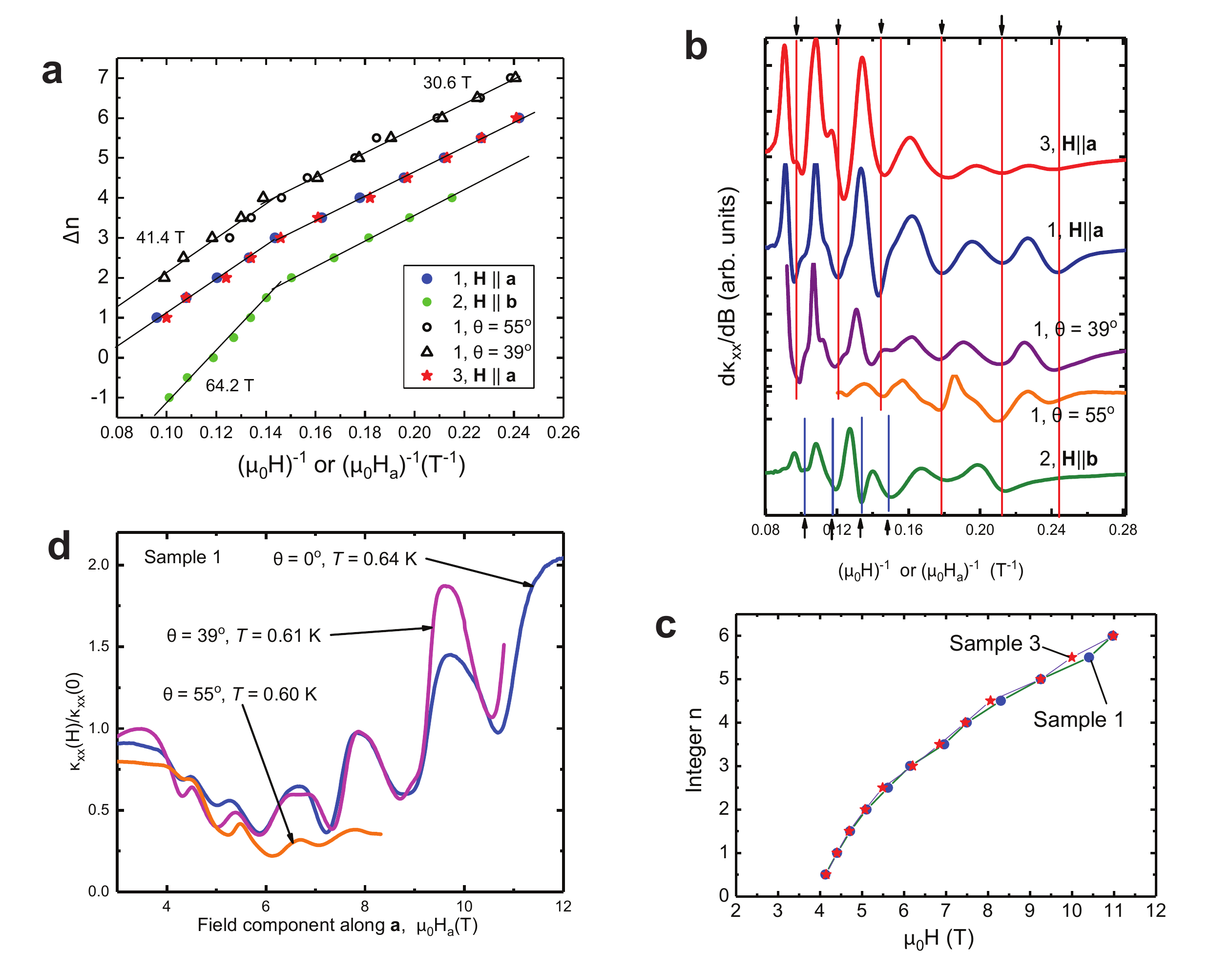}
\caption{\label{fig2} 
}
\end{figure*}

We next show that the oscillations are periodic in $1/H$. Figure \ref{fig2}a displays plots of the integer increment $\Delta n$ vs. $(\mu_0H_n)^{-1}$, where $H_n$ are fields locating extrema of $d\kappa_{xx}/dB$ plotted in Fig. \ref{fig2}b. We focus first on data shown as solid symbols. The data from Samples 1 (blue circles) and 3 (red stars), measured with $\bf H\parallel a$, fall on a curve comprised of straight-line segments separated by a break-in-slope at $\sim$7 T. The slopes $S_f$ of the straight segments are 41.4 T ($H>$7 T) and 30.6 T ($H<$7 T). When $\bf H\parallel b$ (Sample 2, green circles), similar behavior is obtained, with the low-field slope $S_f$ also at 30.6 T. However, the high-field slope is steeper with $S_f$ = 64.2 T. As shown in Figs. \ref{fig1}b and c, the periods are $T$ independent from 0.43 to 4.5 K.

Taken together, the data shown in Figs. \ref{fig2}a to \ref{fig2}d provide strong evidence that the oscillations are intrinsic and reproducible across samples. The five data sets discussed in Panel a were derived from extrema of the derivative curves $d\kappa_{xx}/dB$ displayed in Figs. \ref{fig2}b. The profiles show the close agreement in both period and phase between Samples 1 and 3. The matching of the extrema is especially evident in Fig. \ref{fig2}c, which also shows that periodicity vs. $H$ (as opposed to $1/H$) can be excluded. In Sample 2 the period and phase also agree with 1 and 3 for $H<$ 7 T (the period is shorter above 7 T, as already noted). 

Oscillations observed with $\bf H$ tilted in the $a$-$c$ plane (at an angle $\theta$ to $\bf a$) provide tests in an independent direction. Figure \ref{fig2}c shows curves of $\kappa_{xx}$ measured in Sample 1 with $\theta = 0$ (blue curve), 39$^\circ$ (purple) and 55$^\circ$ (orange) (curves of $\kappa_{xx}$ at various $T$ are in Figs. \ref{figS4}a and \ref{figS4}b). By plotting the curves vs. $H_a = H\cos\theta$, we find that the periods match quite well (with a possible phase shift for the curve at 55$^\circ$). The corresponding derivatives at 39$^\circ$ and 55$^\circ$ are plotted in Panel (b). We infer that, in tilted $\bf H$, the periods depend only on $H_a$. Moreover, the close matching of the blue and purple curves strongly supports an intrinsic origin. 

In  Sec. \ref{Strain} in Methods, we discuss the relation of the oscillations to de Haas van Alphen experiments on the correlated insulator SmB$_6$~\cite{Sebastian}. For the proposed mechanism~\cite{Cooper,FaWang} to apply to $\alpha$-RuCl$_3$, we would need $H$ $\sim$10,000 T. We also elaborate further on the evidence against artifactual origins, such as stacking faults produced by field-induced strain~\cite{Kubota,Cao}.


The profile of $\Delta\kappa_{amp}$ vs. $H$ actually imposes a tight constraint on possible mechanisms. Above 11.5 T in the polarized state, the oscillations vanish abruptly. Below $H_C$ the oscillations survive as a weak tail extending to 4 T in the zig-zag state. The amplitude profile suggests a close connection to the QSL state. The $1/H$ periodicity suggests an intriguing analogy with Shubnikov de Haas oscillations, despite the absence of free carriers. We note that Landau-level oscillations have been predicted in the insulating 2D QSL state with $\bf H$ normal to the plane~\cite{Motrunich,Senthil}. A spinon Fermi surface in the QSL state of $\alpha$-RuCl$_3$ is widely anticipated~\cite{Normand,Trivedi,Fujimoto}. Our finding that $S_f$ is determined by $H_a$ suggests either a fully 3D QSL state (or possibly a different mechanism). Nonetheless, quantization of a spin Fermi surface is currently our leading interpretation.

The amplitude $\Delta\kappa_{amp}$ is much larger in Sample 3 than in 1 (from Figs. \ref{figS3}a and \ref{fig1}d, the peak values are 100 and 16.5 mW/Km, respectively). We have uncovered a correlation with lattice disorder, as estimated from $\kappa^{ph}$ (see Sec. \ref{phonons}). In the QSL state, it is difficult to separate reliably the phonon term $\kappa^{ph}$ from the spin-excitation conductivity $\kappa^s_{xx}$ because of strong spin-phonon coupling (which causes oscillations in both). However, in the polarized state above 11.5 T, $\kappa_{xx}$ is strongly dominated by $\kappa^{ph}$ (the profile becomes $H$ independent). Hence the plateau value of $\kappa_{xx}/T$ measures reliably the lattice disorder. At 1.0 K, $\kappa_{xx}/T$ is much higher in Sample 3 (2.2 W/mK$^2$) than in 1 (0.7 W/mK$^2$). The lower disorder in Sample 3 correlates with a 6-fold increase in the oscillation amplitude.

\begin{figure*}[t]
\includegraphics[width=18 cm]{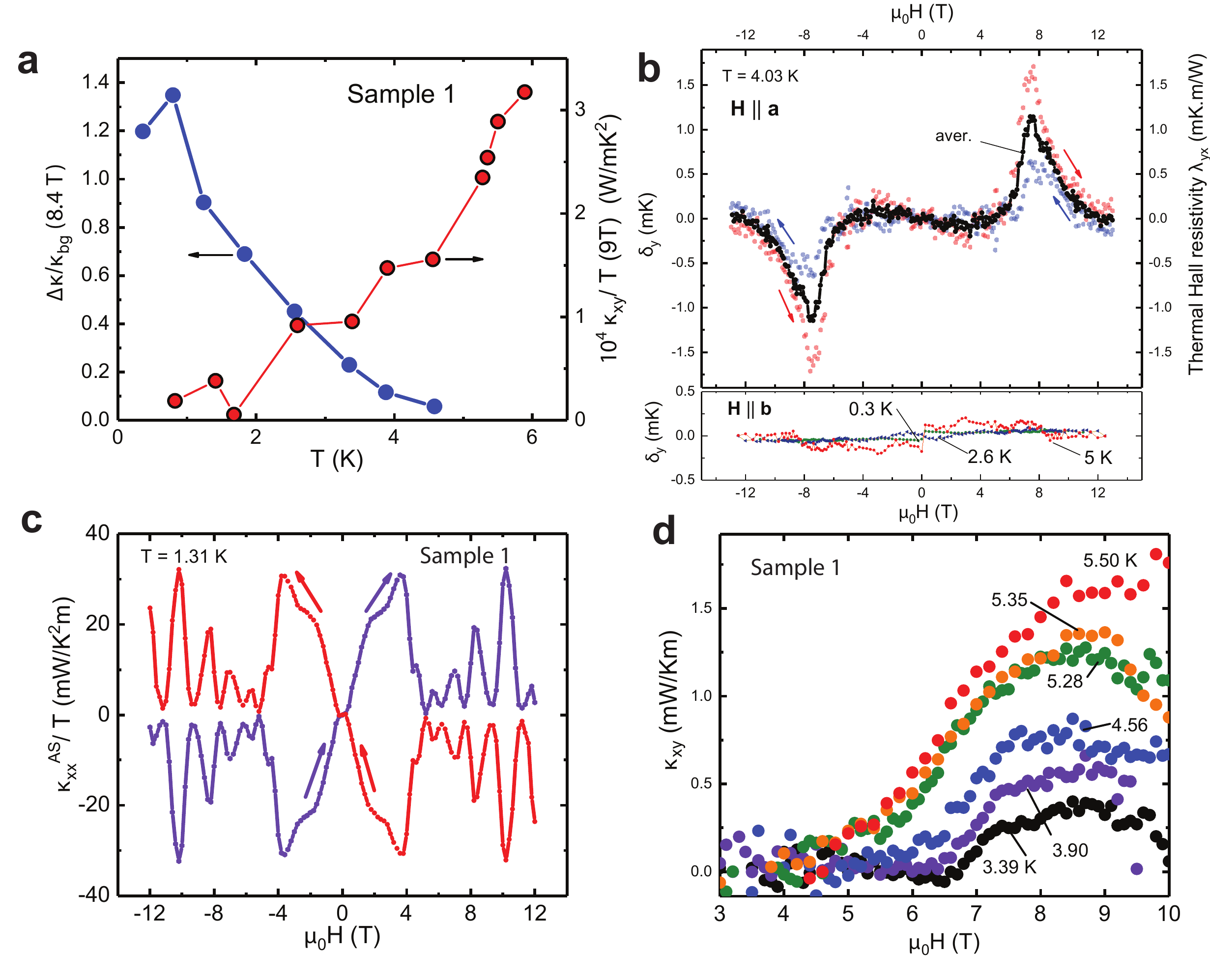}
\caption{\label{fig3} 
}
\end{figure*}

The observed status of $H_a$ seems empirically related to the planar thermal Hall effect (PTHE), which appears only with $\bf H\parallel a$. At a fixed $H$, the ratio $\Delta\kappa/\kappa_{bg}$ decays with $T$ at a rate consistent with an effective mass $m^*\sim 0.64\, m_e$ where $m_e$ is the free electron mass (blue circles in Fig. \ref{fig3}a).  The decay in $\Delta\kappa/\kappa_{bg}$ is accompanied by a rapid growth in the PTHE observed with $\bf H\parallel a$ (red circles). Recently, Yokoi \emph{et al.}~\cite{Yokoi} reported that $\kappa_{xy}/T$ measured with $\bf H\parallel a$ seems to be quantized, within a narrow interval in $T$ (3.8-6 K) and and in $H$ ($10<\mu_0 H<11.2$ T). We have extended the PTHE experiment down to $T$ = 300 mK to gain a broader perspective.

Below 4 K, it is necessary to use the method described in Eqs. \ref{DyT}--\ref{intrinsic} in Methods~\cite{Hirschberger} to isolate the intrinsic thermal Hall signal $\delta_y$ (defined in Eq. \ref{DyT}) from artifacts arising from hystereses in $\kappa_{xx}$ as shown for e.g. in Fig. \ref{fig3}c. For $\bf H\parallel b$, the intrinsic thermal Hall signal is found to be zero for $0<H<14$ T and 0.3$<T<$ 5 K (Fig. \ref{fig3}b, lower panel). However, with $\bf H\parallel a$, a finite $\delta_y$ emerges above $\sim$2 K, as shown in upper panel of Fig. \ref{fig3}b. The antisymmetry of $\delta_y$ with respect to $\bf H$ identifies it as a true PTHE. This is the thermal-conductivity analog of the true planar anomalous Hall effect observed in ZrTe$_5$~\cite{Liang}.

Inverting the matrix $\lambda_{ij}(H)$ to obtain $\kappa_{ij}(H)$, we find that $\kappa_{xy}$ displays a dome profile that grows with $T$ in the QSL phase (Fig. \ref{fig3}d). Together, Figs. \ref{fig3}a and \ref{fig3}d provide a broad view of how the PTHE varies with $T$. While the trends of our $\kappa_{xy}$ are consistent with those in Ref. \cite{Yokoi} (e.g. the PTHE exists only with $\bf H\parallel a$), we note that the strong $T$ dependence evident in Figs. \ref{fig2}d seems difficult to reconcile with a quantized value occuring in the interval 3.8 -- 6 K. Where the two data sets overlap (4-5 K), our magnitudes are much smaller ($\kappa_{xy}\sim$ 0.3 vs. 0.8 mW/K$^2$m at $T$ = 5 K). 

In summary, we have observed quantum oscillations in $\kappa_{xx}$ in $\alpha$-RuCl$_3$ with $\bf H$ in plane. The prominence of the amplitude in the interval (7.3, 11) T implies that they are specific to the QSL state.

\vspace{1cm}
\centerline{* ~~~ * ~~~  *}

\newpage
\vspace{1cm}\noindent
$^{*}$P.C. and T.G. contributed equally to the experiment.\\
$^{\S}$Corresponding author email: npo@princeton.edu\\
$^\dagger$Current address of M.H.: Department of Applied Physics and Quantum-Phase Electronics Center (QPEC), The University of Tokyo, Bunkyo-ku, Tokyo, 113-8656, Japan.\\
$^{\ddagger}$A.B. is also affiliated with the Department of Physics, Purdue University, West Lafayette, IN 47907, USA.

\vspace{5mm}\noindent
{\bf Acknowledgements} \\
We thank Jingjing Lin and Stephan Kim for technical assistance and T. Senthil and I. Sodemann for valuable discussions. The measurements of $\kappa_{xx}$ and P.C. and M.H. were supported by a MRSEC award from the U.S. National Science Foundation (DMR 1420541). T.G. and the low-$T$ thermal Hall experiments were supported by the U.S. Department of Energy (DE-SC0017863). A.B. and S.E.N are supported by the DOE, Office of Science, Scientific User Facilities Division. N.P.O. was supported by the Gordon and Betty Moore Foundation's EPiQS initiative through grant GBMF4539. P.L-K. and D.M. were supported by Moore Foundation's EPiQS initiative through grant GBMF4416.

\vspace{3mm}
\noindent
{\bf Author contributions}\\
P.C. and T.G. performed the measurements and analyzed the data together with N.P.O. who proposed the experiment. M.H. greatly enhanced the experimental technique employed. A.B., P.L-K. and S.E.N. provided guidance on prior results. Crystals were grown at ORNL by P.L-K., J.Y. and D.M. at ORNL. The manuscript was written by N.P.O., P.C. and T.G.

\vspace{3mm}
\noindent
{\bf Additional Information}\\
Supplementary information is available in the online version of the paper.
Correspondence and requests for materials should be addressed to N.P.O.
Reprints and permissions information is available at www.nature.com/reprints.

\vspace{3mm}
\noindent
{\bf Competing financial interests}\\
The authors declare no competing financial interests.

\newpage
\vspace{5mm}
\noindent
{\bf Figure Captions}
\vspace{5mm}\\
\noindent
{\bf Figure 1: Quantum oscillations in the quantum spin liquid (QSL) phase in $\alpha$-RuCl$_3$ (Sample 1).} Panel (a): The phase diagram showing the QSL phase sandwiched between the zig-zag and polarized states with $\bf H\bf \parallel a$ (axes $\bf a$ and $\bf b$ shown in inset). The ZZ2 phase lying between critical fields $B_{c1}$ and $B_{c2}$ is outlined by the dashed curve~\cite{Balz}. Panel (b) shows the emergence of oscillations in $\kappa_{xx}(H)$ ($\bf H\parallel a$) as $T$ falls below 4 K. Data recorded using the stepped-field technique to correct for magnto-caloric effects (Sec. B in Methods). Panel (c) displays the oscillations over the full field range at selected $T$ (data recorded continuously as well as with stepped-field method). At $\sim$11 T, $\kappa_{xx}$ displays a step-increase to a plateau-like profile in the polarized state in which oscillations are strictly absent. The amplitude $\Delta\kappa_{amp}$ (solid circles in Panel (d)) is strikingly prominent in the QSL state. Its profile (shaded orange) distinguishes the QSL from adjacent phases. A weak remnant tail extends below 7 T to 4 T in the zig-zag state. The derivative curves $d(\kappa_{xx}/T)/dB$ show that the oscillations onset abruptly at 4 T. The large derivative peak centered at $\sim$11.3 T corresponds to the step-increase in $\kappa_{xx}$ mentioned, and is not part of the oscillation sequence.

\vspace{3mm}
\noindent
{\bf Figure 2: Periodicity and intrinsic nature of oscillations (panels labelled clockwise).} Panel (a) plots the integer increment $\Delta n$ versus $1/H_{n}$ (or $1/H_{n,a}$) where $H_{n}$ are the fields identifying extrema of the derivative curves $d\kappa_{xx}/dB$ ($H_{n,a} = H_n\cos\theta$ for tilted $\bf H$). Solid symbols represent data taken with $\bf H$ strictly in-plane. The blue circles (Sample 1) and red stars (Sample 3) were measured with $\bf H\parallel a$, whereas the green circles were measured in Sample 2 with $\bf H\parallel b$. Open symbols are measurements in Sample 1 with $\bf H$ tilted in the $a$-$c$ plane at angles $\theta$ = $39^\circ$ (triangles) and $55^\circ$ (circles), relative to $\bf a$. The data sets fall on the same segmented curve (comprised of line segments with slope 31 T below 7 T and 41 T above 7 T). The exception is the high-field slope of 64 T in Sample 2 with $\bf H\parallel b$. For clarity, the 3 data sets are shifted vertically by $\Delta n = 1$.
Panel (b): Curves of the derivative $d\kappa_{xx}/dB$ vs. $1/H$ (or $1/H_a$) for Samples 1, 2 and 3 ($H_a = H\cos\theta$). For Sample 1, we show $d\kappa_{xx}/dB$ measured with $\theta = 0, 39^\circ$ and $55^\circ$. The extrema of $d\kappa_{xx}/dB$ are plotted in Panel (a). Vertical lines mark the values of $1/H_n$ and $1/H_{n,a}$ read off from the straight-line fits in Panel (a) for integer increments $\Delta n$. 
Panel (c): Replot of integer increment $\Delta n$ vs. $H_n$ (fields locating the extrema of $d\kappa_{xx}/dB$) in Samples 1 (blue circles) and 3 (red stars) to check for periodicity vs. $H$ (as opposed to $1/H$). In both data sets (measured with $\bf H\parallel a$), the curve diverges to large negative $\Delta n$ as $H$ decreases to 4 T. The narrowing of the spacing between adjacent extrema is strongly incompatible with periodicity vs. $H$.
Panel (d) shows the effect of tilting $\bf H$ out of the plane in Sample 1 by angle $\theta$ (relative to $\bf a$) at $T\sim$ 0.6 K. The curves are measured with $\theta = 0^\circ$ (blue), $39^\circ$ (purple) and $55^\circ$ (orange). When they are plotted vs. $H_a$, the periods of the oscillations in $\kappa_{xx}$ match well for the 3 angles.

\vspace{3mm}
\noindent
{\bf Figure 3: The planar thermal Hall response.} 
Panel (a)  The $T$ dependence of $\Delta\kappa/\kappa_{bg}$ at 8.4 T (blue circles) and the planar thermal Hall conductivity $\kappa_{xy}$ at 9 T (red). The decrease of $\Delta\kappa/\kappa_{bg}$ with $T$ (consistent with an effective mass $m^*/m_e$ = 0.64) is anti-correlated with the increase in $\kappa_{xy}$. 
Panel (b) shows the emergence of the PTHE signal $\delta_y$ with $\bf H\parallel a$ (upper panel). At $T$ = 4.03 K, $\delta_y$ in Sample 1 (left axis) displays sharp peaks that are antisymmetric in $H$ for $\bf H\parallel a$ (black circles). Corresponding values of $\lambda_{yx}$ are on the right axis. The lower panel shows the null thermal Hall resistivity (expressed as the thermal Hall signal $\delta_y$) measured with $\bf H\parallel b$ at 0.3, 2.6 and 5 K in Sample 2. The total uncertainty in $\delta_y$ is 0.3 mK. 
Panel (c) shows the hysteresis in $\kappa_{xx}$ that can contaminate $\kappa_{xy}$ if not properly subtracted. The right-going (purple) and left-going (red) scans have been antisymmetrized with respect to $H$.
In Panel (d), $\kappa_{xy}(H)$ derived from the measured tensor $\lambda_{ij}$ are plotted for several $T$ from 3.4 to 5.5 K. The dome-shaped profiles are the planar thermal Hall effect reported in Ref. \cite{Yokoi} but in our experiment the values are not quantized.

\newpage

\begin{thebibliography}{10}

\bibitem{Anderson} Anderson, P. W. 
Resonating valence bonds: a new kind of insulator?
\emph{Mater. Res. Bull.} {\bf 8}, 153 (1973).

\bibitem{Ng} Zhou, Y., Kanoda, K. $\&$ Ng, T.K.
Quantum spin liquid states,
\emph{\rmp} {\bf 89}, 025003 (2017).


\bibitem{Savary} Savary, L. $\&$ Balents, L.,
Quantum spin liquids: a review,
\emph{Rep. Prog. Phys.} {\bf 80}, 106502 (2017).



\bibitem{WenQFT} Wen, X.G.,
\emph{Quantum Field Theory of Many-Body Systems},
Oxford Univ. Press (2004), Ch. 9.

\bibitem{Kitaev} Kitaev, A.,
Anyons in an exactly solved model and beyond,
\emph{Ann. Phys.} {\bf 321}, 2-111 (2006).

\bibitem{Plumb}
Plumb, K. W. \etal,
$\alpha$-RuCl$_3$: a spin-orbit assisted Mott Insulator on a honeycomb lattice. 
\emph{Phys. Rev. B} {\bf 90}, 041112 (2014).


\bibitem{Sears1}
Sears, J. A. \etal,
Magnetic order in $\alpha$-RuCl$_3$: A honeycomb-lattice quantum magnet with strong spin-orbit coupling. 
\emph{Phys. Rev. B} {\bf 91}, 144420 (2015).  



\bibitem{Banerjee}Banerjee, A. \etal,
Proximate Kitaev quantum spin liquid behavior in a honeycomb magnet. 
\emph{Nat. Mater.} {\bf 15}, 733-740 (2016). 


\bibitem{Sears2}
Sears, J. A. \etal, 
Phase diagram of $\alpha$-RuCl$_3$ in an in-plane magnetic field. 
\emph{Phys. Rev. B} {\bf 95}, 180411 (2017).

\bibitem{Wang}
Wang, Z. \etal,
Magnetic Excitations and Continuum of a Possibly Field-Induced Quantum Spin Liquid in $\alpha$-RuCl$_3$.
\emph{Phys. Rev. Lett.} {\bf 119}, 227202 (2017).

\bibitem{Leahy}
Leahy. I. A. \etal, 
Anomalous thermal conductivity and magnetic torque response in the honeycomb magnet $\alpha$-RuCl$_3$,
\emph{Phys. Rev. Lett.} {\bf 118}, 187203 (2017). 


\bibitem{Banerjee2}
Banerjee, A. \etal,
Excitations in the field-induced quantum spin liquid state of $\alpha$-RuCl$_3$. npj \emph{Quantum Materials} (2018) 3:8 ; doi:10.1038/s41535-018-0079-2 

\bibitem{Hentrich}
Hentrich, R. \etal, 
Unusual phonon heat transport in $\alpha$-RuCl$_3$: Strong spin-phonon scattering and field-induced spin gap. 
\emph{Phys. Rev. Lett.} {\bf 120}, 117204 (2018).


\bibitem{Lampen1}
Lampen-Kelley , P. \etal,
Anisotropic susceptibilities in the honeycomb Kitaev system $\alpha$-RuCl$_3$. 
\emph{Phys. Rev. B} {\bf 98}, 100403 (2018). 

\bibitem{Lampen2}
Lampen-Kelley, P. \etal,
Field-induced intermediate phase in $\alpha$-RuCl$_3$: Non-coplanar order, phase diagram, and proximate spin liquid,
cond-mat arXiv:1807.06192v1

\bibitem{Balz}
Balz, C. \etal, 
Finite field regime for a quantum spin liquid in $\alpha$-RuCl$_3$. 
\emph{Phys. Rev. B} {\bf 100}, 060405(R) (2019).

\bibitem{Khaliullin} Jackeli, G. $\&$ Khaliullin, G.,
Mott Insulators in the Strong Spin-Orbit Coupling Limit: From Heisenberg to a Quantum Compass and Kitaev Models,
\prl {\bf 102}, 017205 (2009).


\bibitem{Chun}
Chun, S.W. \etal,
Direct evidence for dominant bond-directional interactions in a honeycomb lattice iridate Na$_2$IrO$_3$,
\emph{Nature Physics} {\bf 11}, 462 (2015).

\bibitem{KimKee} Kim, H. S.  $\&$ Kee, H. Y.,
Crystal structure and magnetism in $\alpha$-RuCl$_3$: An \emph{ab initio} study,
\emph{\prb} {\bf 93}, 155143 (2016).


\bibitem{Kasahara} 
Kasahara, Y. \etal,
Majorana quantization and half-integer thermal quantum hall effect in a Kitaev spin liquid.
\emph{Nature} {\bf 559}, 227-231 (2018). 

\bibitem{Sinn}
Sinn, S. \etal,
Electronic Structure of the Kitaev Material $\alpha$-RuCl$_3$ Probed by
Photoemission and Inverse Photoemission Spectroscopies. 
\emph{Sci. Rep.} $\bf 6$, 39544; doi: 10.1038/srep39544 (2016).


\bibitem{Sebastian} Tan, B.S. \etal, 
Unconventional Fermi surface in an insulating state,
\emph{Science} {\bf 349}, 287 (2015). DOI: 10.1126/science.aaa7974

\bibitem{Cooper} Johannes Knolle and Nigel R. Cooper,
Quantum Oscillations without a Fermi Surface and the Anomalous de Haas–van Alphen Effect,
\emph{Phys. Rev. Lett.} {\bf 115}, 146401 (2015). DOI: 10.1103/PhysRevLett.115.146401

\bibitem{FaWang}
Long Zhang, Xue-Yang Song, and Fa Wang,
Quantum Oscillation in Narrow-Gap Topological Insulators,
\emph{Phys. Rev. Lett.} {\bf 116}, 046404 (2016). DOI: 10.1103/PhysRevLett.116.046404


\bibitem{Kubota} Yumi Kubota, Hidekazu Tanaka, Toshio Ono, Yasuo Narumi, and Koichi Kindo,
Successive magnetic phase transitions in $\alpha$-RuCl$_3$: \emph{XY}-like frustrated magnet on the honeycomb lattice,
\emph{\prb} {\bf 91}, 094422 (2015). DOI: 10.1103/PhysRevB.91.094422

\bibitem{Cao} 
H. B. Cao, A. Banerjee, J.-Q. Yan, C. A. Bridges, M. D. Lumsden, D. G. Mandrus, D. A. Tennant, B. C. Chakoumakos, and S. E. Nagler,
Low-temperature crystal and magnetic structure of $\alpha$-RuCl$_3$,
\emph{\prb} {\bf 93}, 134423 (2016).
DOI: 10.1103/PhysRevB.93.134423



\bibitem{Motrunich} Motrunich, O.I.,
Orbital magnetic field effects in spin liquid with spinon Fermi sea: Possible application to $\kappa$-(ET)$_2$Cu$_2$(CN)$_3$,
\emph{\prb} {\bf 73}, 155115 (2006). DOI: 10.1103/PhysRevB.73.155115

\bibitem{Senthil} Sodemann, I., Chowdhury, D. $\&$ Senthil, T., 
Quantum oscillations in insulators with neutral Fermi surfaces,
\emph{\prb} {\bf 97}, 045152 (2018). DOI: 10.1103/PhysRevB.97.045152


\bibitem{Normand} 
Liu, Z. X.  $\&$ Normand, B.,
Dirac and Chiral Quantum Spin Liquids on the Honeycomb Lattice in a Magnetic Field,
\emph{\prl} {\bf 120}, 187201 (2018).



\bibitem{Trivedi} Patel, N. D.  $\&$ Trivedi, N.,
Magnetic field-induced intermediate quantum spiin liquid with a spinon Fermi surface,
\emph{PNAS} {\bf 116}, 12199 (2019).



\bibitem{Fujimoto} Takikawa, D.  $\&$ Fujimoto, S.,
Impact of off-diagonal exchange interactions on the Kitaev spin-liquid state of $\alpha$-RuCl$_3$,
\emph{\prb} {\bf 99}, 224409 (2019).


\bibitem{Yokoi}
Yokoi, T. \etal, 
Half-integer quantized thermal Hall effect in the Kitaev material $\alpha$-RuCl$_3$.
cond-mat arXiv: 2001.01899v1.


\bibitem{Hirschberger} Hirschberger, M. \emph{Quasiparticle excitations with Berry Curvature in insulating magnets and Weyl semimetals} (PhD thesis, Princeton University 2017). 

\bibitem{Liang} Liang, T. \etal., 
Anomalous Hall effect in ZrTe$_5$,
\emph{Nat. Phys.} {\bf 14}, 451-455 (2018). doi.org/10.1038/s41567-018-0078-z



\end{thebibliography}
\end{document}